\documentclass{ws-procs9x6}

\begin{document}

\title{Recent progress on the manipulation of single atoms in optical tweezers for quantum computing}

\author{A. Browaeys$^*$, J. Beugnon, C. Tuchendler, H. Marion, A. Ga\"{e}tan, Y. Miroshnychenko, \\
B. Darqui\'{e}, J. Dingjan,
Y.R.P. Sortais, A.M. Lance, M.P.A. Jones, \\
G. Messin,  P. Grangier}
\address{Laboratoire Charles Fabry de l'Institut d'Optique, CNRS, Univ. Paris-sud,\\
Campus Polytechnique,
RD 128, 91127 Palaiseau cedex, France\\
$^*$E-mail: antoine.browaeys@institutoptique.fr}

\begin{abstract}
This paper summarizes our recent progress towards using single rubidium atoms trapped in an optical tweezer to encode
quantum information. We demonstrate
single qubit rotations on this system and measure the coherence of the qubit. We move the quantum bit over distances of tens of microns and
show that the coherence is preserved. We also transfer a qubit atom between two tweezers and show no loss of coherence.
Finally, we  describe our progress towards conditional entanglement
of two atoms by photon emission and two-photon interferences.
\end{abstract}

\bodymatter

\vspace{1cm}

Quantum computing has been proposed to solve certain classes of computational problems, such as factoring and searching, faster
than using a classical computer\cite{Nielsen}~. In addition, one could engineer these quantum computers
in such a way that they could perform
simulations of  quantum systems.
From a  fundamental point of view, a quantum computer can be thought as a collection of two-level systems,
well isolated from the environment,
which can interact with each other in a controlled way.
Building  such a quantum computer may therefore help to understand decoherence
of a macroscopic quantum system towards a classical system.

The practical implementation of a quantum computer relies on a physical system that
constitutes a good approximation of a two-level system.
Among all the systems proposed so far, neutral atoms present the advantage of
well controlled manipulations of the internal and external degrees
of freedom. Furthermore, neutral atoms offer  built-in scalability when the atoms are trapped in
periodic potentials.

Following this route, we have chosen to encode the quantum information on two hyperfine states of a single rubidium atom
trapped in an optical tweezer. Using this system, it has been demonstrated that several tweezers can be arranged in an array,
with each sites well localized and adressable
by optical methods\cite{Bergamini04,Dumke02}~.

This paper describes  how we trap and observe a single atom in an optical tweezer created by focusing a
far-off resonant laser down to a micrometer size waist. We then show the coherent manipulation of the
two-level system and characterize the coherence of this quantum bit. As a first step towards the controlled interaction
between two atoms
trapped in an array of tweezers, we  demonstrate a scheme
where  the qubit is transfered between two tweezers, with no observed loss of coherence
and no change in the external degrees of freedom of the atom.
Additionnally, we  move the atom over distances that are typical of the separation
between atoms in an array of optical traps, and show that this transport does not affect the coherence of the qubit.
Finally, we are working towards the conditional entanglement of two atoms trapped in tweezers separated by 10 microns.
We have shown two key ingredients of this protocol: the controlled emission of a single photon by a single atom and the
two-photon interference of  photons emitted by two atoms.

\section{Diffraction-limited optics for single-atom manipulation}

Our optical tweezer is a far off-resonance dipole  trap, with a size of about one micrometer. We produce this tweezer by
focusing a laser beam down to the diffraction limit of a large numerical aperture aspherical lens.

We use a lens manufactured by LightPath Technologies with a numerical aperture of 0.5. The working distance between the lens and the focal point is 6 mm,
large enough to allow a good optical
access around the lens. We have separately tested the lens using
a wavefront analyzer. We have found a RMS deviation of the resulting wavefront of less than $\lambda/30$ over the whole
numerical aperture, thus demonstrating that the aspherical lens is  diffraction limited.

\begin{figure}
\begin{center}
\psfig{file=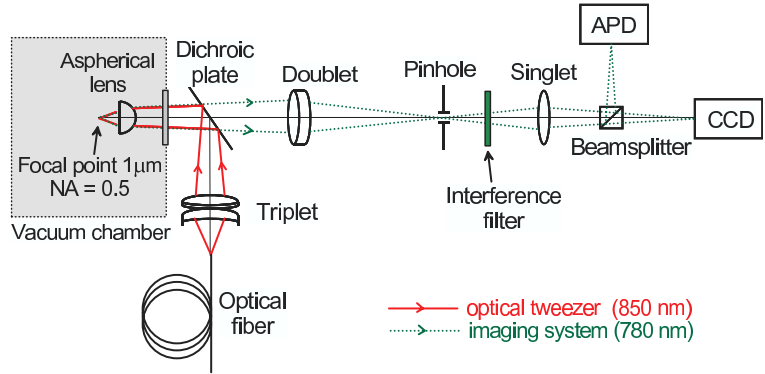,width=10cm}
\end{center}
\caption{Optical setup of the single-atom trapping (solid line) and imaging systems (dotted lines). }
\label{layout_aspherix}
\end{figure}

The optical layout, consisting of the aspherical lens and standard optical elements, is shown in figure~\ref{layout_aspherix}.
The dipole trap laser beam, produced by a 850 nm laser diode, is sent throught a single mode optical fiber and is shaped
at the output of the fiber
by a triplet lens. The beam goes through the viewport of the vacuum chamber inside which the aspherical lens is placed.
Before putting this system together, we have checked on a separated bench that the whole optical system is diffraction limited
(see reference\cite{Sortais07} for more details).

The same lens is used to collect the fluorescence light (780 nm) emitted by the atom trapped in the tweezer. This fluorescence
light is collected outside the vacuum chamber using a confocal imaging system, as represented in figure~\ref{layout_aspherix}.
The light is sent onto an avalanche photodiode, used in counting mode, and a CCD camera.
This imaging system is also diffraction limited and has a spatial resolution of 1 micrometer\cite{Sortais07}~.

\begin{figure}
\begin{center}
\psfig{file=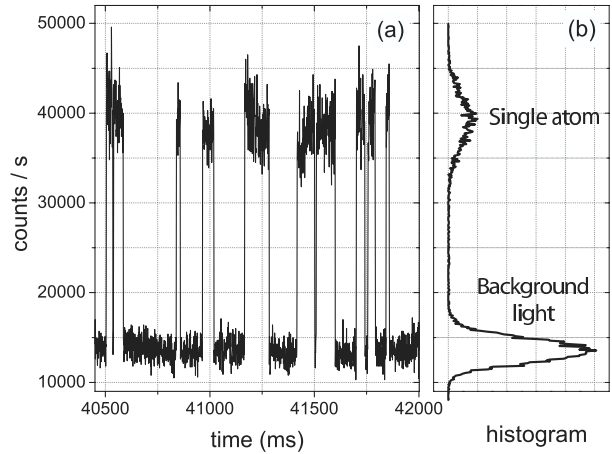,width=8cm}
\end{center}
\caption{(a) Fluorescence of a single atom measured by the avalanche photodiode. (b) Histogram of the measured fluorescence
recorded over 100 sec. The two well separated peaks in this histogram correspond to the absence of atom and
the presence of exactly one atom.}
\label{fluo_single_atom}
\end{figure}

We  use this optical system to trap and observe single atoms in the tweezer.  The tweezer
is loaded from the cloud of rubidium atoms cooled in an optical molasses.
Figure~\ref{fluo_single_atom} shows an example of the signal obtained on the avalanche photodiode versus time.
The  steps in the signal
correspond to the fluorescence of a trapped atom at 780 nm, induced by the molasses laser.
The absence of double steps is an indication that only individual atoms are trapped. This single-atom
trapping is made
possible by a ``blockade mechanism'' studied in detail in reference\cite{Schlosser02}~.

\section{Single-atom quantum bit}

Our quantum bit is encoded on the $|0\rangle = |F =1, M=0\rangle$,
$|1\rangle = |F =2, M=0\rangle$ hyperfine sublevels of a rubidium 87 atom.
This choice of levels for the qubit provides the advantage of zero first-order sensitivity to  magnetic
fields.

We initialize the qubit in state $|0\rangle$ by optical pumping, with an efficiency of 85 \%.
We read the state of the qubit using a state selective measurement limited by the quantum projection noise.
For this purpose, we send a laser on resonance with state $|1\rangle$ and the state $F'=3$ connected by a transition
at 780 nm. Radiation pressure expels the atom out of the trap, if the atom is initially in state $|1\rangle$.
Otherwise, if the atom is in state $|0\rangle$, the laser
leaves the atom unaffected. We then check for the presence of the atom. This method therefore maps the internal state of the atom
on the presence or the absence of the atom at the end of the sequence.

\begin{figure}
\begin{center}
\psfig{file=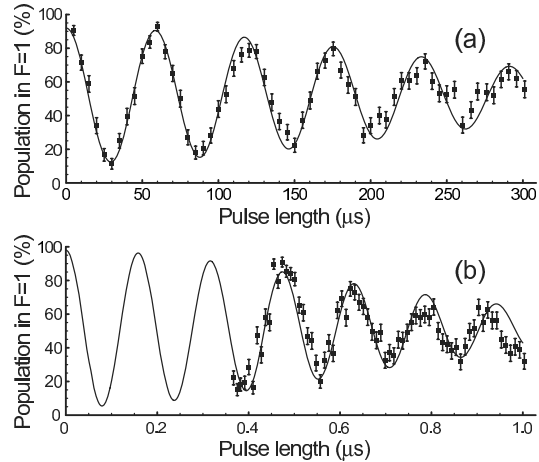,width=7cm}
\end{center}
\caption{Single-atom Rabi oscillations. Fraction of atoms in $F=1$, as a function of the Raman pulse length, at low (a) and
high (b) intensity. Details can be found in reference\cite{jones07}~. }
\label{Rabi}
\end{figure}

We drive the qubit transition with two Raman lasers, one of them being the dipole trap. The two beams are colinear and sent through
the same optical fiber and the large numerical aperture lens. Due to the tight focusing of the two lasers, we observe a Rabi frequency
of the two-photon transition as high as $2 \pi \times 6.7$ MHz. Figure~\ref{Rabi} shows the population of state $|0\rangle$
for two durations of the Raman pulse\cite{jones07}~.

We study the coherence of the quantum bit using Ramsey spectroscopy. We apply a first $\pi/2$
Raman pulse to prepare the atomic state $(|0\rangle+|1\rangle)/\sqrt{2}$. We let the system evolve
and we apply a second $\pi/2$ pulse. The signal exhibits oscillations at a frequency given by the detuning of the
Raman lasers with respect to the qubit transition. The decay of the contrast of the oscillations as a function of the time interval
between the two pulses is the signature of the loss of coherence.  We attribute this decay to the residual motion of the atom in the trap, together
with the fact that the two states of the qubit experience a slightly different trapping potential, leading to a dephasing of the quantum bit. Our best
 $1/e$ dephasing time is $ 630\ \mu$sec. For details, see reference~\cite{jones07}~.

\begin{figure}
\begin{center}
\psfig{file=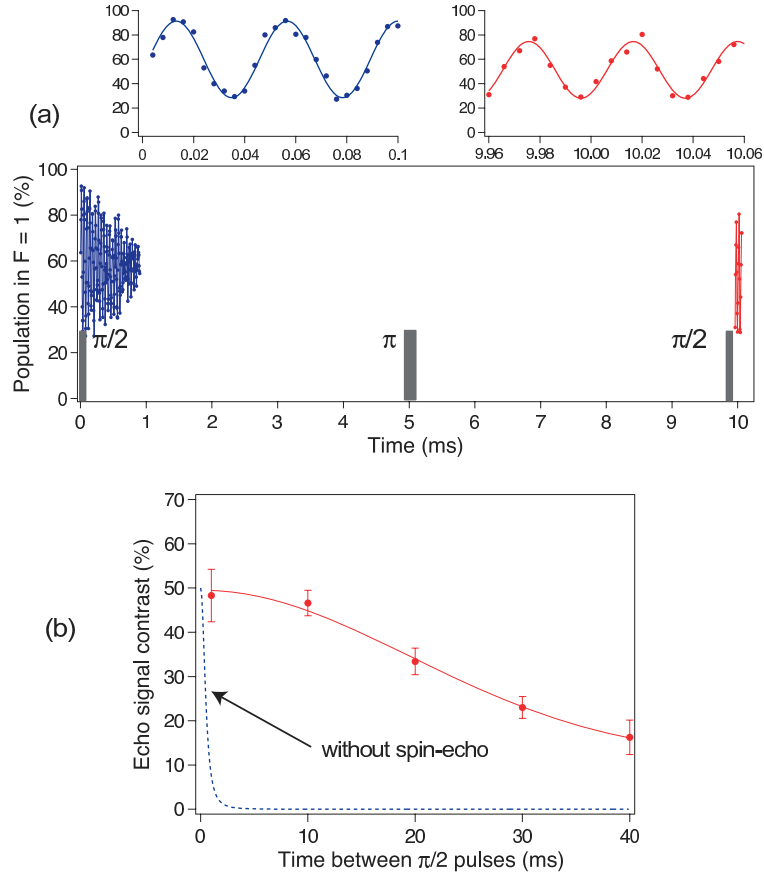,width=10cm}
\end{center}
\caption{Example of the spin-echo signal. Figure (a) shows the revival of the oscillations after the $\pi$
pulse has been applied. Figure (b) shows the amplitude of the echo signal for different durations of the spin-echo sequence.}
\label{spinecho}
\end{figure}

We rephase this dephasing  by inserting a $\pi$ pulse between the two $\pi/2$ pulses. Using this spin-echo technique,
we observe a revival of the oscillations
after a time  as long as 40 ms, corresponding to a 70 fold improvement with respect to the coherence time of the qubit, as shown in
figure~\ref{spinecho}.

\section{Transport and transfer of  atomic qubits}

Neutral atoms are promising candidates for the realization of a large-scale quantum register. To perform a quantum
computation, a key feature is the ability to perform a gate between two arbitrary qubits of the register.
As a first step, we have demonstrated a scheme where the  atomic qubit is
transfered between two  tweezers,
and then transported over several tens of micrometers.
We show that these manipulations of the external degrees of
freedom preserve the coherence of the qubit, and do not induce any heating.
The distance travelled is typical of the separation
between atoms in an array of dipole traps. These techniques can also be useful to position an atom at the
node of the electromagnetic
field in an optical cavity for QED experiments\cite{Nussmann05}~.

To show that the transfer preserves the coherence of the quantum bit, we prepare a superposition
$(|0\rangle + |1\rangle)/\sqrt{2}$ in a first tweezer using a first $\pi/2$ pulse.
We then decrease the depth of the first tweezer while increasing the
depth of a second tweezer superimposed with the first one. After a dwelling time of 200 $\mu$sec, we transfer the atom back to the
initial tweezer and apply a second $\pi/2$ pulse. We vary the time between the two pulses around 200 $\mu$sec
and observe the corresponding
Ramsey oscillations. We have shown that the amplitude of the Ramsey signal remains unchanged whatever the depth of the second trap
is. We have also checked that when the depths of the two traps are identical, the ``temperature'' of the atom is unchanged.

\begin{figure}
\begin{center}
\psfig{file=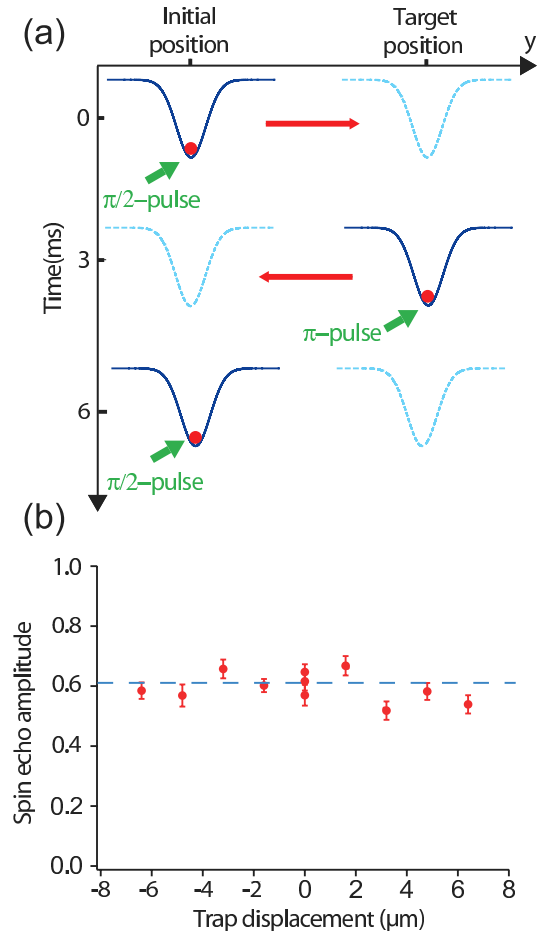,width=7cm}
\end{center}
\caption{(a) Principle of the moving qubit experiment, with the position of the tweezer when the various pulses
of the sequence are applied. (b) Amplitude of spin-echo signal versus the amplitude of the displacement. }
\label{movingqubit}
\end{figure}

We  have also moved the tweezer when the qubit is prepared in a superposition  $(|0\rangle + |1\rangle)/\sqrt{2}$.
We have transported the atom up to $\pm 9$ $\mu$m away from the axis of the focusing lens, and brought it back to its
initial position\cite{beugnon07}~. The entire round trip takes 6 ms. As this time is longer than the dephasing time of the quantum bit, we apply
a $\pi$ pulse when the tweezer is at its turning point, as shown in figure~\ref{movingqubit}.
We have measured that the amplitude of the spin-echo signal remains constant when we varied the amplitude of the displacement.
This  indicates that the coherence of the quantum bit is preserved during the transport.

Finally, we have measured the phase of the Ramsey sequence during the transfer experiment and the phase of the spin-echo signal
during the transport. We have modelled this phase evolution and found a good agreement with the result of the experiment. This
understanding of the phase is crucial for a possible implementation in a quantum computer where qubit phases need to be controlled.

\section{Towards conditional entanglement of two atoms}

The ability to generate entanglement is a key feature that any physical system should exhibit to be useful for any quantum information
processing task. For example,  entanglement is essential in teleportation protocols between two parties. Entanglement is also a necessary ingredient
in the two approaches to quantum computing. In the quantum circuit approach\cite{Nielsen}~, entanglement is
generated during the course of the implementation of the algorithms. In the cluster state approach\cite{Raussendorf01}~, it even constitutes the starting point of
the calculation. It is therefore important to be able to generate and control the
entanglement in this quantum system.

\begin{figure}
\begin{center}
\psfig{file=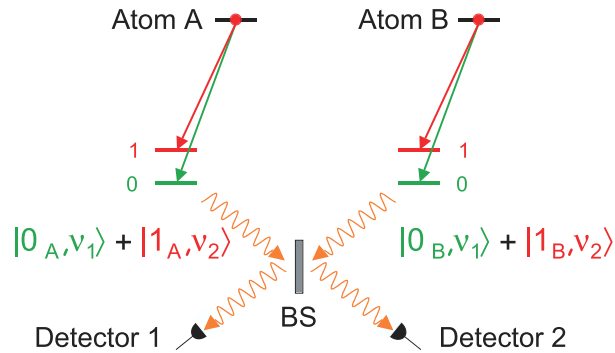,width=8cm}
\end{center}
\caption{Principle of a conditional entanglement of two atoms $A$ and $B$ based on the entanglement
between each atom and an emitted photon followed by the recombination of the two photons on a 50/50 beam-splitter (BS).}
\label{conditional_entanglement}
\end{figure}

Among all the methods proposed to entangle two neutral atoms, conditional entanglement based on photon emission
is promising, as it does not require any direct interaction between the atoms. A simplified scheme is presented in
figure~\ref{conditional_entanglement} (see for example references\cite{simon03,Moehring07}).
We isolate three levels in the atom. The upper level is connected to the two logic levels
by an optical transition, with frequencies $\nu_{1}$ and $\nu_{2}$. Each atom can decay to level $|0\rangle$ and $|1\rangle$ by emitting a photon with equal probability.
The scheme entangles the internal state of the atom with the frequency of the emitted photon, generating for atom $A$ the state
$(|0_{A},\nu_{1}\rangle +|1_{A},\nu_{2}\rangle)/\sqrt{2}$. The state of the two photons and two atoms $A$ and $B$
is therefore
\begin{equation*}
|\psi\rangle = \frac{1}{\sqrt{2}} (|0_{A},\nu_{1}\rangle +|1_{A},\nu_{2}\rangle) \otimes
\frac{1}{\sqrt{2}}(|0_{B},\nu_{1}\rangle +|1_{B},\nu_{2}\rangle)\ .
\end{equation*}
Re-arranging the terms, this state can be re-written as
\begin{eqnarray*}
|\psi\rangle  =  \frac{1}{2} (& &|0_{A},0_{B}\rangle \otimes |\nu_{1},\nu_{1}\rangle +\\
& &|1_{A},1_{B}\rangle \otimes |\nu_{2},\nu_{2}\rangle +\\
& &\frac{1}{\sqrt{2}}(|0_{A},1_{B}\rangle + |1_{A},0_{B}\rangle)\otimes \frac{1}{\sqrt{2}}(|\nu_{1},\nu_{2}\rangle + |\nu_{2},\nu_{1}\rangle)+ \\
& &\frac{1}{\sqrt{2}}(|0_{A},1_{B}\rangle - |1_{A},0_{B}\rangle) \otimes \frac{1}{\sqrt{2}}(|\nu_{1},\nu_{2}\rangle - |\nu_{2},\nu_{1}\rangle)\ )\ .
\end{eqnarray*}
When the two photons are recombined on a beam-splitter, a two-photon interference prevents coincidences on the  two detectors
when the two-photon state before the beam-splitter is $|\nu_{1},\nu_{1}\rangle$ and $|\nu_{2},\nu_{2}\rangle$.
The absence of simultaneous coincidence is also
true when the two-photon state is
$|\nu_{1},\nu_{2}\rangle + |\nu_{2},\nu_{1}\rangle$, as the two-photon amplitudes of each component exactly cancel out
(see for example reference\cite{Moehring07}).
Therefore a simultaneous detection event
on the two detectors heralds the preparation
of the entangled atomic state $(|0_{A},1_{B}\rangle - |1_{A},0_{B}\rangle)/\sqrt{2}$. In this scheme, two-photon interferences acts as
a ``filter'', and the preparation is heralded by the double detection.

Three ingredients are required to implement this protocol. The first one is the triggered emission of a single photon by a single
atom. The second one is the observation of two photon interferences, and the third one is the ability to entangle an atom with
an emitted photon.  In the following sections we describe our implementation of the two first steps. We note that the third step has been
realized by two groups in the recent years\cite{Blinov04,Volz06}~.

\section{Single atom as a single-photon source}

\begin{figure}
\begin{center}
\psfig{file=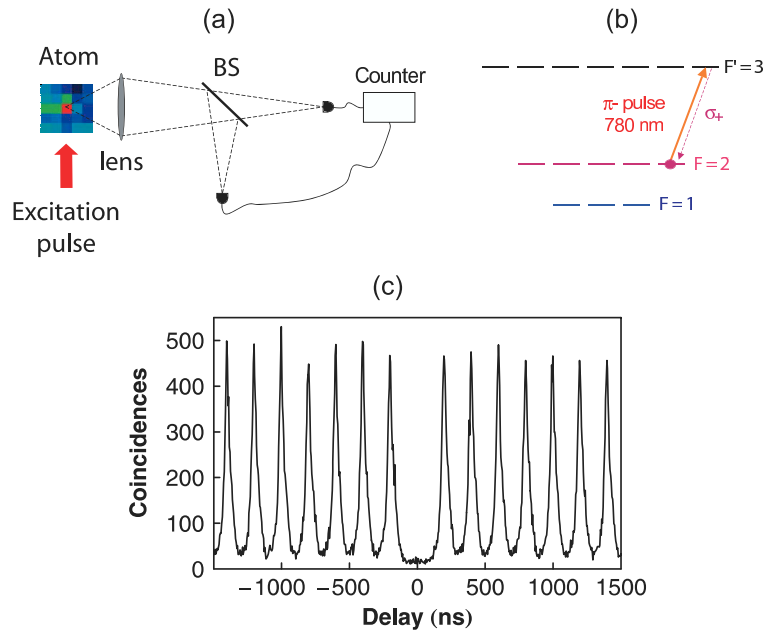,width=10cm}
\end{center}
\caption{(a) Principle of the single-photon source based on an atom trapped at the focal point of a large numerical aperture lens.
(b) Relevant hyperfine levels  of rubidium 87 used in the experiment. (c) Histogram of the coincidences measured on the two
single-photon counters. }
\label{single_photon_source}
\end{figure}

We control the emission of single photons by a single atom placed at the focal point of a large numerical aperture lens by sending
$\pi$ pulse on the optical transition connecting $(F=2,M=2)$ and $(F'=3, M'=3)$, see figure~\ref{single_photon_source}.
The duration of the pulses is 4 ns, and
the separation between the pulses is 200 ns. The
emitted photons all have the same $\sigma^+$ polarization.

We collect
0.6 \% of the emitted photons on an avalanche photodiode.  We have characterized the single-photon nature of the source by
measuring the coincidences on two photodetectors placed in the imaging system~\cite{Darquie05}~. The resulting curve is shown in
figure~\ref{single_photon_source}(c). The absence of coincidence at zero delay is the signature that single photons are emitted by the atom.
A careful analysis of this curve shows that the probability that the source emits two photons following the same excitation pulse is
1.8\%. This number is in good agreement with a calculation taking into account the 4 ns duration of the excitation pulse, which is not
negligible with respect to the 26 ns lifetime of the upper state.

\section{Interference of two photons emitted by two atoms}

When two indistinguishable single photons are fed into the
two input ports of a beam-splitter, the photons will
leave together from the same output port. This is a quantum
interference effect, which occurs because the two possible paths,
where the photons leave in different output ports, interfere
destructively. This effect was first observed in parametric
downconversion by Hong, Ou and Mandel\cite{Hong87}~.

\begin{figure}
\begin{center}
\psfig{file=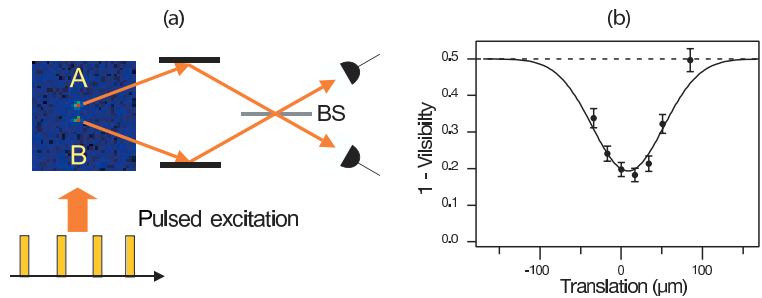,width=10cm}
\end{center}
\caption{(a) Principle of the two-photon interference experiment. The photons emitted by two atoms $A$ and $B$ are
recombined on a 50-50 beam-splitter (BS), and the coincidences are measured using two single-photon detectors
(b) Influence of the wavefront matching. We plot the
amplitude of the residual coincidence signal at zero delay for various relative distance between the two photon modes, translated
parallel to each other. For perfect interference, the signal should go all the way down to zero when the two modes are not translated.  }
\label{two-photon_interference}
\end{figure}

We have shown the interference of two photons emitted by two atoms trapped in two tweezers separated by 6 microns  (see details
in reference~\cite{beugnon06}).
The two atoms are excited by the same 2 ns-laser pulse following the procedure described in the previous
section. We recombine the two photons of same polarization on an optical setup equivalent to
a 50/50 beam-splitter and we measure the
coincidences at zero delay on two photodiodes placed in the outpout ports of the beam-splitter, as represented in
figure~\ref{two-photon_interference}(a). If the two-photon interference were perfect,
one should not measure any coincidence as the two photons must leave in the same output port.

To analyze the visibility of the interferences, we varied the spatial overlap between the photons propagating in free space.
In order to do so, we translated the spatial mode of one photon with respect to the other one. The result is represented in
figure~\ref{two-photon_interference}(b). For our best overlap, we find a visibility of the interferences of 60\%, coming
from the difficulty to mode-match the two photons. This number is a measure of the indistinguishability of the two
interfering photons. This mode-matching can be improved by coupling the photons to single mode fibers.

\section{Conclusion}

We have demonstrated some basic manipulations of a single quantum bit encoded on an
atom trapped in an optical tweezer. We have shown
internal state rotation at the single atom level and we have measured the internal coherence of the qubit state.
We have also demonstrated two necessary ingredients of a conditional entanglement protocol. Our current estimate gives
an efficiency to produce one entangled pair on the order of $10^{-6} - 10^{-5}$. We anticipate a rate  of entangled pair
production of one every 100 seconds.
\vspace{5mm}

{\bf Acknowledgments:}
We acknowledge financial support from
IFRAF, ARDA/DTO and the European Integrated project SCALA. LCFIO is  CNRS UMR8501.
M.P.A. Jones and A.M. Lance are supported by Marie Curie Fellowships. A. Ga\"{e}tan is supported by a DGA Fellowship.


\vspace{5mm}

\begin{thebibliography}{30}

\bibitem{Nielsen} M.A. Nielsen, \& I.L. Chuang {\it Quantum Computation and Quantum Information}.
Cambridge University Press (2000).

\bibitem{Bergamini04} S. Bergamini, {\it et al.\/},   {\it J. Opt. Soc. Am. B} {\bf 21}, 1889 (2004).

\bibitem{Dumke02} R.  Dumke,  {\it et al.},
{\it Phys. Rev. Lett.} {\bf 89}, 097903 (2002).

\bibitem{Sortais07} Y.R.P.  Sortais, {\it et al.},
{\it Phys. Rev. A} {\bf 75}, 013406 (2007).

\bibitem{Schlosser02} N. Schlosser, G. Reymond and P.  Grangier,
{\it Phys. Rev. Lett.} {\bf 89}, 023005 (2002).

\bibitem{jones07} M.P.A. Jones, J. Beugnon, A. Ga\"{e}tan, J. Zhang, G. Messin, A. Browaeys, P. Grangier, {\it Phys. Rev. A}
{\bf 75}, 013406(R) (2007).

\bibitem{Nussmann05} S. Nussmann, M. Hijlkema, B. Weber, F. Rohde, G. Rempe, and A. Kuhn,
{\it Phys. Rev. Lett.} {\bf 95}, 173602 (2005).

\bibitem{beugnon07} J. Beugnon, {\it et al.\/}, in press {\it Nature Physics}  (2007); arXiv:0705.0312 [quant-ph].

\bibitem{Raussendorf01} R. Raussendorf and H. Briegel, {\it Phys. Rev. Lett.} {\bf 86}, 5188 (2001).

\bibitem{simon03} C. Simon, W.T.M. Irvine, {\it Phys. Rev. Lett.\/} {\bf
91}, 110405 (2003).

\bibitem{Moehring07} D. L. Moehring, M. J. Madsen, K. C. Younge, R. N. Kohn, Jr.,
P. Maunz, L.-M. Duan,  C. Monroe and B.B. Blinov, {\it J. Opt. Soc. Am. B} {\bf 24}, 300 (2007).

\bibitem{Blinov04} B.B. Blinov, D.L. Moehring, L.-M. Duan, C. Monroe, {\it
Nature\/} {\bf 428}, 153 (2004).

\bibitem{Volz06} J. Volz, M. Weber, D. Schlenk, W. Rosenfeld, J. Vrana,  K. Saucke,
C. Kurtsiefer and H. Weinfurter, {\it Phys. Rev. Lett.} {\bf 96}, 030404 (2006).

\bibitem{Darquie05} B.  Darqui\'e, {\it et al.\/},  {\it Science} {\bf 309},
454-456 (2005).

\bibitem{Hong87} C.K. Hong,  Z.Y. Ou, and L. Mandel,  {\it Phys. Rev. Lett.} {\bf 59}, 2044-2046 (1987).

\bibitem{beugnon06} J. Beugnon, {\it et al.\/}, {\it Nature} {\bf 440},
779 - 782 (2006).
\end{thebibliography}
\end{document}